\documentstyle{mn}
\input{epsf}
\newcommand{\be}{\begin{eqnarray}}
\newcommand{\ee}{\end{eqnarray}}
\newcommand{\hMpc}{{\ifmmode{h^{-1}{\rm Mpc}}
\else{$h^{-1}$Mpc}\fi}}

\title[Statistical significance of LQG]
{Testing statistical significance of large quasar groups with sheets model of large scale structure}
\author[Pilipenko, Malinovski]
{S.V. Pilipenko$^{1,2}$, A.M. Malinovsky$^1$\\
$1$Astro Space centre of Lebedev Physical
           Institute of  Russian Academy of Sciences, Profsojuznaja st. 84/32,
                        117997 Moscow,  Russia\\
        $2$ Moscow Institute of Physics and Technology, Institutskij per. 9, 
            141700 Dolgoprudnyj, Russia\\}

\begin{document}
\maketitle
\begin{abstract}
We argue that the largest group of quasars (LQG) U1.27 discovered by Clowes et al. (2013) in the SDSS DR7 catalogue does not contradict the hypothesis of Poisson distribution of quasars. We found that random catalogues with the same shape and number of QSOs as the real sample may contain groups which resemble U1.27. By simulating quasar catalogues with embedded model of the large scale structure we also found that the size of LQGs selected by MST and similar methods does not correspond to the scale of homogeneity of the Universe and can be explained by the percolation process.
\end{abstract}

\section{Introduction}
Large Quasar Groups (LQG) are known to be the largest structures seen in the Universe at redshifts $z\sim1-2$. LQGs have sizes 40 -- 350 Mpc and typically consist of 5 -- 40 members. A review of the discovery and study of LQGs is given in Clowes et al. (2012).

The LQGs are usually identified with the methods similar to ``friends-of-friends'': a quasar is connected to a cluster if the distance between this quasar and any quasar from the cluster is lower than some threshold length. Only clusters with masses higher than some minimal mass are called LQGs Hereafter we call the number of members of a group its mass. Such methods have an advantage that clusters of arbitrarily shape can be identified.

Recently the discovery a LQG of size $\sim$ 1240 Mpc with membership of 73 quasars has been reported by Clowes et al. (2013). In this Paper we test two conclusions of Clowes et al. (2013): the high statistical significance of this group and its incompatibility with the scale of homogeneity for the concordance cosmology.
\section{The Method}
Following Clowes et al. (2013) we select the sample of quasars from DR7QSO catalogue (Schnider et al., 2010). The magnitude cut $i\leq 19.1$ and the redshift cut $0.8<z<1.9$ have been applied. We use the Minimal Spanning Tree (MST) method for clustering analysis as described in Pilipenko (2007). The code used to calculate the MST is written by V. Turchaninov. We also use the Qhull code (http://www.qhull.org) to compute convex hulls and reproduce the Convex Hull of Member Spheres (CHMS) method of volume estimation used by Clowes et al. (2012, 2013).

Another sample is selected from the DR8 using the SDSS Catalog Archive Server. This sample is constructed with the same redshift and magnitude limits as the previous one. The main difference between DR7 and DR8 is that all quasar data in SDSS has been reprocessed with a new pipeline. As the consequence, 1238 QSOs from DR7-based sample don't exist in DR8. Similarly, 725 QSQs from  DR8-based sample cannot be matched in DR7. Some redshifts of quasars which exist in both samples has changed and distances of about 2000 quasars differ by more than 10 Mpc in two samples.

For the comparison with the DR7QSO-based sample we generate a set of random catalogues. For this purpose we conserve the celestial coordinates of objects and randomly assign them distances in a manner which holds the number density of objects constant with redshift. We use the ``Luxury Pseudorandom Number'' generator.

We also propose a method of generating catalogues which contain a model of the large scale structure. The method is inspired by the Zel'dovich pancake model (Zel'dovich 1970) and by the model of Buryak\& Doroshkevich (1996). For doing so we distribute in a 3D space a set of random plane round sheets of radius $R$. The sheets are randomly oriented and their number density is selected such as to ensure that the mean separation between them along a random line equals $D$. The sheets are populated by randomly distributed points and the mean number density in the catalogue is $n$. The points are shifted in direction perpendicular to sheets by random Gaussian-distributed distances with dispersion $q$.
\section{Results and Discussion}
Firstly we examine the DR7QSO data by searching for the clusters with threshold mass $M=73$ and varying threshold linking length $\ell$. We indeed find the U1.27 group at the smallest $\ell=96$ Mpc.

Clowes et al. (2013) estimate the statistical significance of the group by the CHMS method (Clowes et al. 2012). However in this method the significance does not depend on the volume of the sample in which the group was identified. This is important since in an arbitrarily large volume a group of any kind can be found.
We check the probability of finding a similar group in a sample similar to DR7QSO sample by comparing it with $10^4$ random catalogues. The groups with $\ell=96$ Mpc and $M=73$ were identified in 144 catalogues what gives the probability of U1.27 group to be random 1.4\% and the statistical significance 2.45$\sigma$, which is substantially lower than 3.81$\sigma$ found by Clowes et al. (2013).

The properties of U1.27 are consistent with the properties of random clusters. The main properties: RMS sizes along principal axes 2L, 2H, 2W, the tree and trunk length $L_{tree}$, $L_{trunk}$, mean edge length $\langle \ell \rangle$ and CHMS volume (not bias-corrected) are listed in Table 1. Since clusters in random catalogues can have $M>73$, the CHMS volume was normalized by cluster mass: $V_{CHMS}=73V/M$. It is clear from this table that all quantities are within $1-2\sigma$ from values for random clusters.

\begin{figure}
\centering
\epsfbox{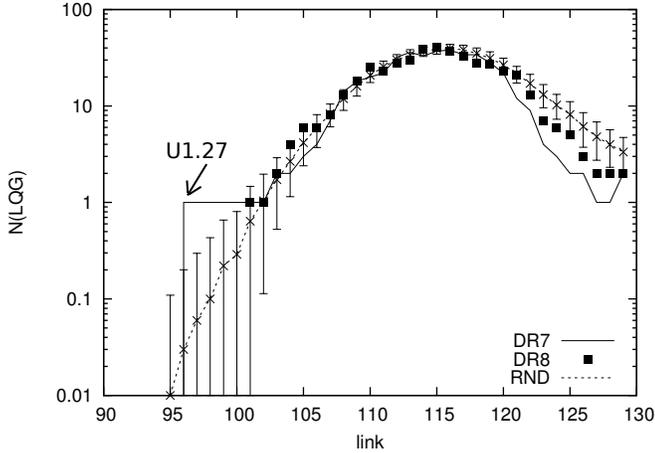}
\caption{The number of LQGs with mass $M\geq73$ as a function of threshold linking length. The solid line represents the DR7-based sample, the dashed line with errorbars represents random catalogues and filled squares represent DR8-based sample.}
\label{fig_nlqg}
\end{figure}

The probability of 1.4\% does not allow to certainly distinguish whether this group is a random coincidence or a signature of the large scale structure. In order to do this a more detailed analysis is required. We suggest that the LSS should manifest itself in some other ways. A detailed comparison of real and random samples is shown in Figure \ref{fig_nlqg} where the number of LQGs as a function of the MST threshold link is plotted. It is clear from this figure that U1.27 group is an outlier and there are no other manifestations of structure.

Another test of significance is the search for this group in the DR8 catalogue. As is clear from Figure \ref{fig_nlqg}, there are no significant deviations from the Poisson distribution in the DR8-based sample. The first group with parameters comparable to U1.27 is detected only at $\ell=101$ Mpc. However it does not coinside with the U1.27 group. The most of quasars from the U1.27 group appear only at $\ell=105$ Mpc when 4 groups are detected.



We compare in detail the U1.27 group in both DR7 and DR8-based samples. All 73 QSOs exist in both samples, however the redshifts of QSOs are slightly different: $\langle \Delta z\rangle=-0.0013$, $\sigma_{\Delta z}=0.0030$. The largest change is $\Delta z=-0.01$. The corresponding difference in QSO distances: $\langle \Delta r\rangle=-2.8$ Mpc, $\sigma_{\Delta r}=6.6$ Mpc, the largest $\Delta r=-20.8$ Mpc. So the significance of the group strongly depends on the algorithm used to measure redshifts. The spectra of quasars in the DR7QSO catalogue have been examined manually by Schnider et al. (2010). However, since the significance of the group strongly depends on the small errors in redshifts, a more careful visual inspection of the spectra of all U1.27 quasars is required to make the final conclusions.

From these results it seems that the properties of the U1.27 LQG are consistent with the Poisson distribution of quasars. We explain this consistency by the fact that DR7QSO catalogue is not very deep. The quasars are too sparse to trace the large scale structure of the Universe which is rather week at redshift $z\sim1$ and scales $\sim100$ Mpc.

\begin{table}
\caption{Comparison of properties of U1.27 group and clusters from random catalogues. Random values are shown with 1$\sigma$ errors.}
\label{tab1}
\begin{tabular}{|l|l|l|}
\hline
Property & U1.27 & random \\
\hline
2L, Mpc & 553 & $520\pm120$ \\
2W, Mpc & 284 & $300\pm40$ \\
2H, Mpc & 167 & $196\pm37$ \\
L/H & 3.31 & $2.80\pm0.98$ \\
W/H & 1.70 & $1.59\pm0.43$ \\
$L_{tree}$, Mpc & 4750 & $5700\pm620$ \\
$L_{trunk}$, Mpc & 2390 & $2420\pm400$ \\
$L_{trunk}/L_{tree}$ & 0.50 & $0.43\pm0.07$ \\
$\langle \ell \rangle$, Mpc & 66 & $69\pm2$ \\
$V_{CHMS}$, Mpc$^3$ & $1.02\cdot 10^8$ & $(1.04\pm0.20)\cdot 10^8$ \\
\hline
\end{tabular}
\end{table}

We are also interested in the following question: what kind of structures may future next-generation deep surveys reveal? To address this question we use the sheets model described in Section 2. The typical distance between large walls in the standard $\Lambda$CDM model is $D\approx 85$ Mpc (Doroshkevich et al. 2003; Demianski\& Doroshkevich 2004;  Semenov 2013). We take for simplicity $D=R=2q$ and find that the statistical significance of large groups depends on the number density of quasars.

The number density of SDSS quasars in the DR7-based sample is $n\approx 4\cdot10^{-7}$ Mpc$^{-3}$. We find that for this $n$ the predicted number of LQGs with $M>73$ in our model and in Poisson catalogues differs only slightly. In particular, for $\ell=96$ Mpc the probability of finding a LQG in a Poisson catalogue is 1.4\% while in a catalogue with sheets this probability increases to only 2.8\% (which corresponds to 2.2$\sigma$ statistical significance).

By contrast, when the number density is 10 times higher, $n=4\cdot10^{-6}$ Mpc$^{-3}$, our model predicts the existence of several LQGs in the SDSS volume with $M>700$, statistical significance $>7\sigma$ (in comparison with Poisson catalogues) and maximal sizes $>15D=1275$ Mpc.
These clusters contain points which inhabit different sheets, however the MST and similar methods ``travel'' from one sheet to another through the places where they intersect. This is an example of the well known percolation process. The threshold link of percolation can be computed for the Poisson distribution of points (Shandarin \& Zeldovich 1986), however for the catalogue which has some underlying structure it should be smaller, which will lead to the increased number of LQGs in comparison with the Poisson distribution. In other words, for a catalogue with strong LSS the plot of the number of LQGs as a function of threshold link is shifted left in comparison with that for a random catalogue.

This means that clusters with sizes of $>$1000 Mpc may be typical in surveys like BOSS, however this size is not connected with the scale of the homogeneity of the Universe.
\section{Conclusions}
We found that the probability to find a LQG similar to U1.27 in a random catalogue is 1.4\% (2.45$\sigma$) or even 2.8\% (2.2$\sigma$) if we take into account the large-scale structure of the Universe modelled by plane sheets. Both these quantities are significantly higher than the probability estimated by Clowes et al. (2013).

We also point out that MST-like algorithms tend to find large clusters with high significance by travelling through the multiconnected large scale structure. The sizes of such clusters do not correspond directly to the size of scale of homogeneity of the Universe.

After the submission of the first variant of
this paper, Dr. Seshadri Nadathur kindly drew our attention to his
recent paper (Nadathur 2013), in which he came to similar conclusions. However, he
obtained slightly different numerical values, possibly due to the use of
the different approach to the problem.

\end{document}